# Active optics in astronomy – Modeling of deformable substrates : Freeform surfaces for FIREBall and MESSIER

**Gerard R. Lemaitre**

Laboratoire d'Astrophysique Marseille - LAM, Aix Marseille Université - AMU

38 rue Frédéric Joliot-Curie, 13388 Marseille CX 13, FRANCE

E-mail: gerard.lemaitre@lam.fr

**Abstract:** Active optics techniques on large telescopes and astronomical instrumentations provide high imaging quality. For ground-based astronomy, the co-addition of adaptive optics again increases angular resolution up to provide diffraction-limited imaging at least in the infrared. Active and adaptive optics marked milestone progress in the detection of exoplanets, super-massive black holes, and large scale structure of galaxies. This paper is dedicated to highly deformable active optics that can generate non-axisymmetric aspheric surfaces - or freeform surfaces - by use of a minimum number of actuators: a single uniform load acts over the surface of a vase-form substrate whilst under reaction to its elliptical perimeter ring.

Two such instruments are presented, 1) the FIREBall telescope and MOS where the freeform reflective diffraction grating is generated by replication of a deformable master grating, and 2) the MESSIER wide-field low-central-obstruction TMA telescope proposal where the freeform mirror is generated by stress figuring and elastic relaxation. Freeform surfaces were obtained by plane super-polishing. Preliminary analysis required use of the optics theory of 3rd-order aberrations and elasticity theory of thin elliptical plates. Final cross-optimizations were carried out with Zemax raytracing code and Nastran FEA elasticity code in order to determine geometry of the deformable substrates.

## 1 Introduction

Rapid technological advances in astronomical instrumentation during the second part of the twentieth century gave rise to 4m, 8m and 10m class telescopes that were completed with close loop *active optics* control in monolithic mirrors (R. Wilson, 1996 [1]), segmented primary mirrors (J. Nelson et al., 1980 [2]), and segmented in-situ stressed active optics mirrors (Su D.-q., X. Cui et al., 2004 [3]). Beside new telescopes launched into space, further advances on ground-based telescopes used *adaptive optics* for blurring the image degradation due to atmosphere. Both *active and adaptive optics* provided milestone progress in the detection of super-massive black holes, exoplanets, and large-scale structure of galaxies.

This paper is dedicated to *highly deformable active optics* that can generate non-axisymmetric aspheric surfaces - or *freeform surfaces* - by use of a minimum number of actuators : a single uniform load acts over the monolithic surface of a vase-form substrate whilst under reaction to its elliptical perimeter ring. The freeform surfaces are the basic aspheric components of dispersive/reflective Schmidt systems. They are either a reflective diffraction grating or a mirror (G. R. Lemaitre, 2009 [4]).

Two such instruments with freeform surfaces are presented :

1) the FIREBall instrument is a NASA/CNES balloon-borne experiment that studies the faint diffuse intergalactic medium from emission lines in the ultraviolet window around 200 nm at 37 km flight altitude. The FIREBall experiment is a second-generation instrument and has been launch in September 2018. It uses a 1-m telescope fed with two-mirror relay and a four-mirror MOS where the freeform reflective diffraction grating is replicated from a deformable master grating [4], (R. Grange et al., 2014)[5], and,

2) the MESSIER wide-field low-central-obstruction three-mirror-anastigmat (TMA) telescope proposal dedicated to the survey of extended astronomical objects with extremely low surface brightness. The optical design leads to a high image quality without any diffracting spider. This prototype is intended to serve as a fast-track pathfinder for a future space-based MESSIER mission. The elliptical freeform mirror is generated by stress figuring and elastic relaxation technique [4], (E. Muslimov et al., 2018)[6].

Freeform surfaces were obtained by plane super-polishing. Preliminary analysis required use of the optics theory of 3rd-order aberrations and elasticity theory of thin elliptical plates. The final cross-optimizations were carried out with Zemax ray-tracing code and `Nastran` FEA elasticity code providing accurate determination of the deformable substrate geometries.

## 2   Optical design with a reflective Schmidt concept

Because of the Schmidt wide-field capability that only requires a basic two-mirror anastigmat, with one free from correcting all three primary aberrations, such a system has been widely investigated as well as for telescope or spectrograph designs. The corrector mirror or reflective diffraction grating is always located at the center of curvature of a spherical concave mirror.

Compared to a Schmidt with a refractor-correcting element, which is a centered-system, a reflective Schmidt must avoid any central obstructions and then necessarily requires a tilt of the optics. The inclination of the mirrors then forms a *non-centered-system*. For an $f/2.5$ focal-ratio, the tilt angle is typically of about 10° and somewhat depending on the FoV size.

For the reflective FIREBall spectrograph, with a collimator at $f/2.5$ and camera mirror at $f/2.5$, each concave mirror of the MOS requires a folding flat mirror.

For the reflective MESSIER telescope proposal at $f/2.5$, a supplementary folding flat mirror was found necessary to avoid any spider in the beam, thus leading to a TMA design.

An important optical feature have be studied and discussed to define shape of the freeform surface of a non-centered system. This surface is with elliptical symmetry. The freeform surface provides the balance of the quadratic terms with respect to the bi-quadratic terms.

Let assume a two-mirror non-centered system where the primary mirror M1 is a freeform and the secondary mirror M2 is a concave spherical surface of radius of curvature $R$. The input beams are circular collimated beams merging at various field angles of a telescope and define along the M1 freeform mirror an elliptical pupil due to inclination angle $i$. A convenient value of the inclination angle allow the M2 mirror to avoid any obstruction and provide focusing very closely to mid-distance between M1 and M2, then very near the distance $R/2$ from M1 (Fig. 1).

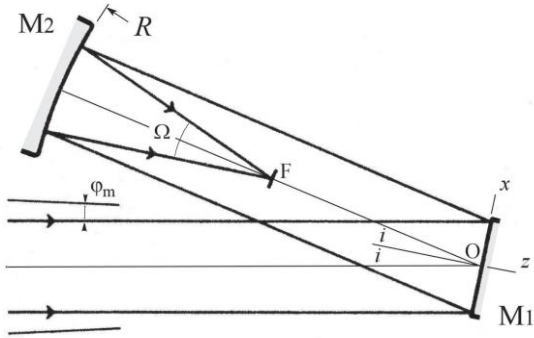

**Fig. 1.** Schematic of a reflective two-mirror Schmidt telescope. The center of curvature of the spherical mirror M2 is located at the vertex of the plane-aspheric freeform mirror M1. The circular incident beams have circular cross-sections with input pupil located at mirror M1. Deviation $2i$ occurs at the principal rays. The semi-field maximum angle – here assumed circular – is denoted $\varphi_m$.

Let us denote $x_m$ and $y_m$ the semi-axes of the elliptic clear-aperture on M1 primary mirror. If the $y$-axis is perpendicular to the symmetry plane $x, z$ of the two-mirror telescope, one defines

$$\Omega = R/4y_m \qquad (1)$$

a dimensionless ratio $\Omega$, where $y_m$ is the semi-clear-aperture of the beams in $y$-direction, i.e. half-pupil size in $y$-direction. The *focal-ratio* of the two-mirror telescope is denoted $f/\Omega$.

It can be shown that the shape $Z_{Opt}$ of the M1 freeform mirror is expressed in first approximation by [4]

$$Z_{Opt} \simeq \frac{s}{\cos i} \left[ \frac{3}{2^7 \Omega^2 R}(h^2 x^2 + y^2) - \frac{1}{8R^3}(h^2 x^2 + y^2)^2 \right], \text{ with } h^2 = \cos^2 i, \qquad (2)$$

and where the dimensionless coefficient $s$ could be used as under-correction parameter slightly smaller than unity ($0.990 < s < 1$). In fact, for a non-centered system the coefficient $s$ must just be set to $s = 1$. A preliminary analysis of the two-mirror system shows that for four equidistant points of a circular field of view, the largest blur image occurs at the largest deviation angle of the field along the $x$-axis. In $y$-direction sideways blur images have an averaged size.

The length from the vertex O of the M1 mirror to the focus F can be derived as [4]

$$OF = \frac{1}{2}\left(1 + \frac{3}{2^6 \Omega^2}\right) R \qquad (3)$$

showing that this distance is slightly larger than the Gaussian distance $R/2$.

The optimal size of blur images for an $f/4$ two-mirror anastigmat telescope over the field of view are displayed by Fig. 2.

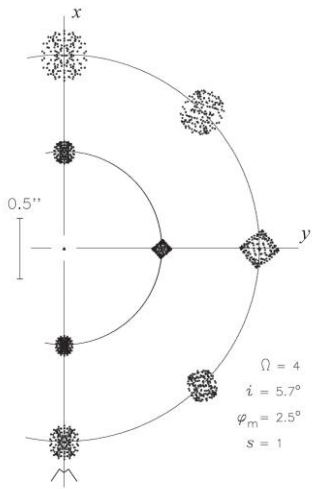

**Fig. 2.** Best residual aberrations of an $f/4$ two-mirror reflective Schmidt ($\Omega = 4$) in the *non-centered system* form. Inclination angle $i = 5.7°$. Semi-field of view $\varphi_m = 2.5°$. The under-correction parameter is just set to $s = 1$. The radius of curvature of the spherical focal surface is also given by $R_{FoV} = OF$ in Eq. (3). The largest blur image corresponds to that with highest deviation of the FoV.

Preliminarily analyses also show that the M1 mirror shape has opposite signs between quadratic and bi-quadratic terms. The shape is given by Eq. (2) and represented by Fig. 3 in either directions $x$ or $y$ in dimensionless coordinates of $\rho$.

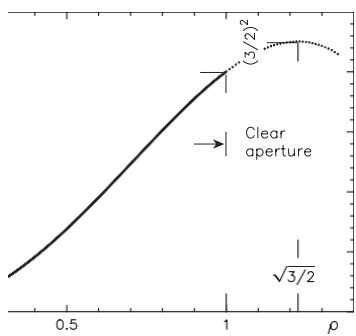

**Fig. 3.** The freeform primary mirror, of biaxial symmetry, is generated by homothetic ellipses having principal lengths in a cos $i$-ratio. One shows that whatever $x$ and $y$ directions the null-power zone is outside the M1 clear-aperture, and in a geometrical ratio $\rho_0 = \sqrt{3/2}\rho_{max} \simeq 1.224\rho_{max}$

Dimensionless coordinates $z$, $\rho$ have been normalized for a clear semi-aperture $\rho = 1$, presently in the $y$-direction, with a maximum sag

$$z(1) = 3\rho^2 - \rho^4 = 2, \qquad (4)$$

which leads to algebraically opposite curvatures $d^2z/d\rho^2$ for $\rho = 0$ and $\rho = 1$.

The conclusions from the best optical design of an all-reflective two-mirror Schmidt telescope with *optimal angular resolution* are the followings (Lemaitre, 1979, 2009)[7][4] :

→ **1.** *For a circular incident beam, the elliptical clear-aperture of the primary mirror is smaller by a ratio $\sqrt{3/2}$ times smaller to that of the null-power zone ellipse.*

→ **2**. *The angular resolution $d_{NC}$ of a reflective non-centered two-mirror telescope is*

$$d_{NC} = \frac{3}{256\Omega^3}\left(\frac{3}{2}i + \varphi_m\right)\varphi_m. \qquad (5)$$

→ **3**. *The primary mirror of a non-centered two-mirror system provides the algebraic balance of second derivative extremals. Therefore, its central curvature is opposite to the local curvatures at edge.*

## 3  Elasticity design and deformable primary-mirror substrate

Active optics preliminarily analysis of a Schmidt's primary mirror M1 leads to investigate deformable substrates where the optical surface is generated by homothetic ellipses.

Let consider the system coordinates of an elliptical plate and denote *n* the normal to the contour *C* of the surface. Equation of *C* is represented by (Fig. 4)

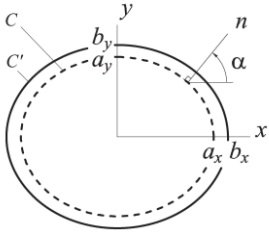

$$\frac{x^2}{a_x^2} + \frac{y^2}{a_y^2} - 1 = 0. \qquad (6)$$

**Fig. 4.** Top view of an *elliptical vase form*. The clear aperture of M1 mirror is included into a smaller surface to that of a constant thickness plate delimited by elliptical contour *C* (dotted line) and defined by semi-axe radii ($a_x$, $a_y$).
An outer ring is built-in to the plate at contour *C* where *n*-directions are normal to *C*. The outer ring is delimited by an homothetic elliptic contour *C'* defined by semi-axes radii ($b_x$, $b_y$).

The bi-laplacian equation of the flexure where a uniform load *q* is applied to the inner plate is

$$\nabla^4 z \equiv \partial^4 z/\partial x^4 + 2\partial^4 z/\partial x^2 \partial y^2 + \partial^4 z/\partial y^4 = q/D, \qquad (7)$$

where the rigidity *D* is

$$D = Et^3/[12(1-v^2)], \qquad (8)$$

*E* the Young's modulus, *v* the Poisson ratio, and *t* the constant thickness of the plate over *C*.

Remaining within the optics theory of third-order aberrations, such optical freeform surfaces can be obtained from elastic bending by mean of the following conditions:
– a flat constant-thickness plate, $t = constant$,
– a uniform load *q* applied all over the surface substrate,
– and a link at the edge to an elliptic contour expressed by a *built-in edge*, or a *clamped edge*, i.e. where the slope is null all along the contour *C* of the plate.

Assuming that the three conditions below are satisfied, the analytic theory of thin plates allows deriving a biquadratic flexure $Z_{Elas}(x, y)$ in the form (Timoshenko & Woinovsky-Krieger)[8], (Lemaitre, 2009)[4],

$$Z_{Elas} = z_0 \left(1 - \frac{x^2}{a_x^2} - \frac{y^2}{a_y^2}\right)^2, \qquad (9)$$

where $z_0$ is the sag at origin and where $a_x$ and $a_y$ are semi-axes corresponding to the elliptic null-power zone of the principal directions of contour *C*. The flexural sag $z_0$ is obtained from substitution of Eq.(9) in biharmonic Eq.(7)

$$z_0 = \frac{q}{8D} \frac{a_x^4 a_y^4}{3a_x^4 + 2a_x^2 a_y^2 + 3a_y^4}. \qquad (10)$$

The null-power zone contour *C* is larger by a factor $\sqrt{3/2}$ to that of the clear-aperture (cf. Fig. 3). The dimensions of semi-axes $a_s$ and $a_y$ at *C* of the built-in ellipse – or null-power zone – relatively to that of clear-semi-apertures $x_m$ and $y_m$ are

$$a_x^2 = 3x_m^2/2 = 3y_m^2/2 \cos^2 i \quad and \quad a_y^2 = 3y_m^2/2. \qquad (11)$$

From Eqs.(2) and (1), in setting $s = 1$ and $x = 0$, and for the built-in radius $y = \sqrt{3/2} y_m$ of the vase form, we obtain the amplitude of the flexure $z_0$ in Eq.(9), as

$$z_0 = \frac{9 y_m}{2^{11} \Omega^3 \cos i}. \qquad (12)$$

From Eqs.(10),(11) and (12) we obtain, after simplification, the thickness *t* of the inner plate

$$t = 8\Omega \left[\frac{2(1-v^2 \cos i)}{3(3+2\cos^2 i + 3\cos^4 i)} \frac{q}{E}\right]^{1/3} y_m. \qquad (13)$$

This defines the execution conditions and elasticity parameters of an elliptical plate where the clear aperture of primary mirror M1 uses a somewhat smaller area than the total built-in surface as delimited by ellipse *C*.

The conclusions from a best optical design of the primary mirror M1 of a two-mirror Schmidt telescope, as corresponding to the profile displayed by Fig. 3, are as follow (Lemaitre, 2009)[4] :

→ **1**. *A built-in elliptic vase-form is useful to obtain easily the primary mirror of a two-mirror anastigmat.*

→ **2**. *The elliptic null-power zone at the plate built-in contour is $\sqrt{3/2}$ times larger than that of the elliptic clear-aperture of the primary mirror.*

→ **3**. *The total sag of the built-in contour is 9/8 times larger than that of its optical clear-aperture.*

The optimization of a built-in substrate, of course, requires losing some of the outer surface which then is not usable by an amount of 33% outside the clear-aperture area. It is clear that a flat deformable M1 substrate, conjugated with an elliptic inner contour, provides most interesting advantages in practice with a *built-in condition* (Fig.5).

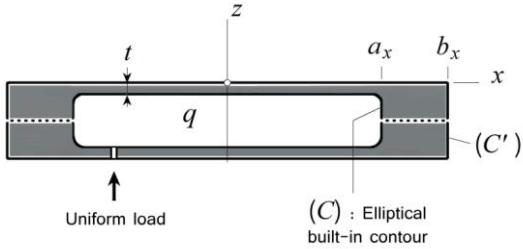

**Fig. 5.** Section of an *elliptical closed vase form*. This form is made two vase-forms oppositely jointed together in a built-in link. The uniform load $q$ is applied inside the deformable substrate over the elliptical contour $C$ of semi-axe radii $(a_x, a_y)$. Outer contour $C'$ can be made circular if $(b_x, b_y)$ are sufficiently larger than semi-axe radii of $C$

An extremely rigid outer ring designed for a *closed vase form*, while polished flat at rest, allows the primary mirror to generate by uniform loading a bi-symmetric optical surface made of *homothetic ellipses* (Fig. 6).

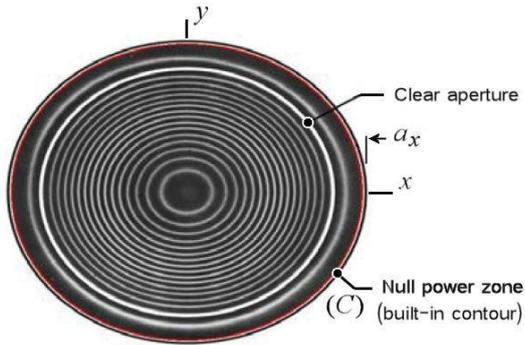

**Fig. 6.** Iso-level lines of a primary mirror M1 generating homothetical ellipses from perfect built-in condition at contour $C$. Elliptic dimensions of the null-power zone over those of clear-aperture must be in a ratio $\sqrt{3/2}$. The algebraic balance of meridian curvatures are then achieved at center and at clear-aperture

A perfect *built-in condition* was applied to the FIREBall design of the freeform bi-quadratic surface, as generated by active bending of a metal stainless steel substrate stress by inner uniform loading. The useful optical surface was a passive replica of a deformable substrate on which a reflective diffraction grating was deposited (cf. Section 4).

Nevertheless some of the outside loosen area of the freeform surface – that did not perfectly fulfilled the theoretical shape of the optical aperture – has been widely reduced by a *closed-form substrate* with a radially *thinned outer elliptic cylinder*, as for MESSIER optimized freeform surface where the pupil shape provides non-significant error from that of theoretical angular resolution (cf. Section 5).

# 4   FIREBall and MOS freeform surface

FIREBall, which stands for Faint Intergalactic Redshifted Emission Balloon, is a NASA/CNES balloon-borne experiment to study the faint diffuse circum-galactic emission in Lyman's ultraviolet line. The field of view of the 1 meter diameter parabola is enlarged using a two-mirror field corrector providing 1000 arcmin$^2$ at the slit mask of a spectrograph (Grange et al., 2014)[5], (Lemaitre et al. 2014)[9].

The Multi Object Spectrograph (MOS) is based on two identical Schmidt systems sharing a reflective aspherical grating. The aspherization of the grating is achieved using a double replication technique of a metallic deformable matrix. We present hereafter the f/2.5 spectrograph design and the deformable matrix process to obtain the freeform Schmidt grating made of homothetic ellipses.

## 4.1   FIREBall basical design and ray-tracing modeling

FIREBall-I and FIREBall-II proposals concern the study of emission lines in the ultraviolet window around 200 nm at 37 km flight altitude. For the previous 2009 flight launched from Fort Sumner (NM), FIREBall-I relied on a fiber bundle Integral Field Unit (IFU) spectrograph fed by a 1 meter diameter parabola.

As the science goals are concentrating on the circum-galactic medium, FIREBall-II (Fig. 7) will use a Multi Object Spectrograph (MOS) for the 2018 flight. It will take full advantage of the new high QE, low noise 13.5 M-pixels UV CCD developed by Caltech/JPL. This will increase the number of targets per flight while keeping the fast f-number of $f/2.5$ to maintain a high signal to noise ratio. Compared to the 2009 flight, the new MOS will have a much larger field of view (400 arcmin$^2$) than the IFU (16 arcmin$^2$) and the image quality of the spectrograph has to match the small pixel size (13.5 $\mu$m) of the new CCD compared to the 60 $\mu$m FWHM of previous photon counting detector. The new goal for the MOS is to obtain a spectral resolution 0.1 nm, R=2200, over a narrow band 200-210 nm while the spatial image quality is 1.5 arcsec FWHM.

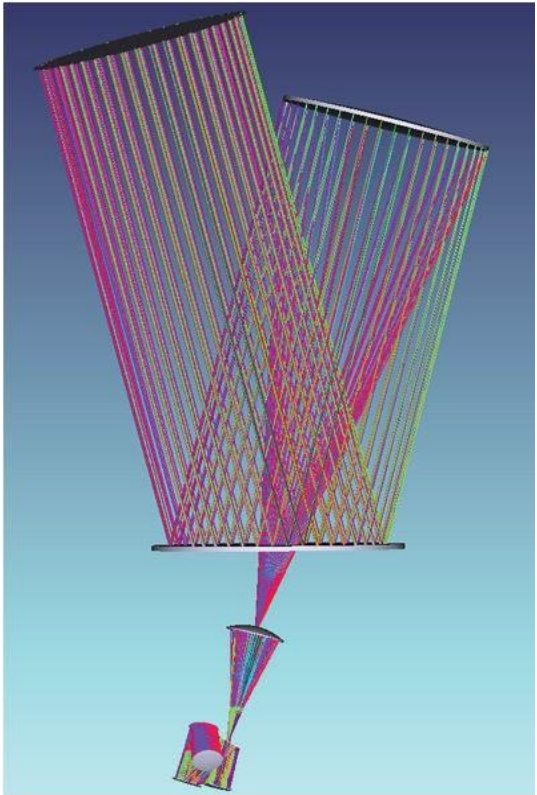

We secondly present the optical design of the field corrector based on two conicoid mirrors working at magnification unity, M=-1, to maintains the $f/2.5$ ratio value (Fig. 8).

Thirdly, the MOS is a double-Schmidt design with both $f/2.5$ spherical mirrors as collimator and camera mirror (Fig. 9). The two-mirror field corrector images the aberrated paraboloid telescope focus onto curved slit masks optimized to produce a flat field of view at the spectrograph focal plane (Table 1 and Fig. 10).

**Fig. 7.** Schematic of FIREBall optical train.

**1.** A flat siderostat mirror, 1,3 m aperture, is followed by a 1 m paraboloid mirror at f/2.5 mounted and stabilized in the gravity.

**2.** A two-mirror image-transport, of magnification M = -1, made of a first convex mirror, provides a curved anastigmatic field of view at the two-mirror transport system.

**3.** The curved field is then imaged by a double-reflective Schmidt MOS $f/2.5$-$f/2.5$ onto a flat CCD.

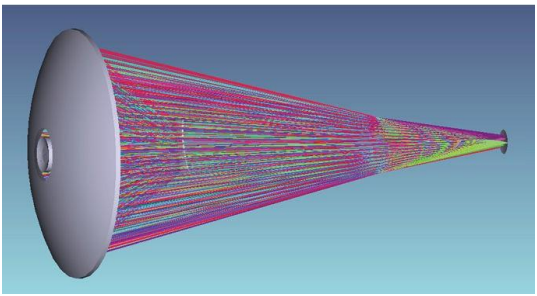

**Fig. 8.** Two-mirror image-transport, magnification M = -1, as field corrector with predetermined curvature. Both mirrors are conicoids where the first mirror is convex. This is a field corrector system where the concentric two-mirror pair has been modified to provide a predetermined curvature with anastigmatic properties that includes the 1 m paraboloid.

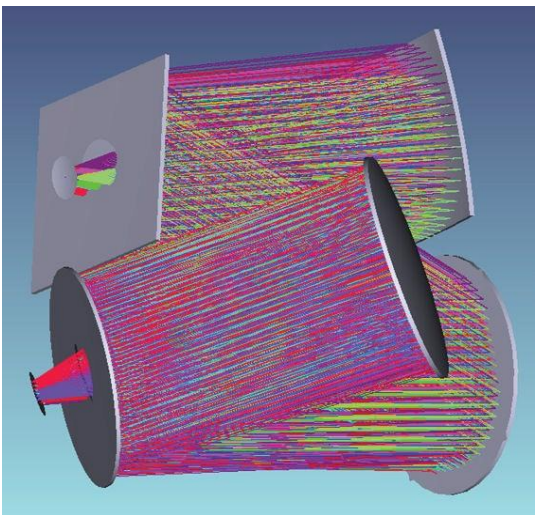

**Fig. 9.** Double Schmidt spectrograph design folded by two flats. Beyond the mask the new spectrograph is based on two identical Schmidt systems acting as collimator and camera both sharing a 2400 ℓ/mm reflective Schmidt grating.

| FIREBall-2 performance Optical review 2030904 updated 20130903 | reference pure schmidt | |
|---|---|---|
| | rms 1 D arcsec | FWHM (2D=1D) arcsec |
| rotation noise (') rms (1 direction) | 3.0 | |
| Max Dist. to guide star (') - at field edge | 30.9 | |
| Max rms 1D noise from rotation (") - rot averaged over image | 0.8 | 1.9 |
| X and Y axis noise (") | 0.9 | 2.1 |
| Jitter X,Y, Theta (") at max dist | 1.1 | 2.5 |
| Telescope figuring (") | 1.0 | 2.5 |
| FieldCorr.+Telescope design (") | 0.6 | 1.3 |
| perfect -> lab -> flight (") | 0.2 | 0.5 |
| pixel sampling | 0.3 | 0.7 |
| Image at mask level  (") | 1.6 | 3.9 |
| on-sky 50 um slit width arcsec= 4.2 arcsec | | |
| on sky resolution orth slit | 2.0 | 4.8 |
| slit width = 0.87 A | | |
| Rayleigh separation | 0.95 | |
| Spectrograph PSF  um | 6.4 | 15.0 |
| Spectrograph PSF(A) | 0.11 | 0.3 |
| Spectro PSF arcsec | 0.53 | 1.3 |
| Sky at detector  along slit " | 1.7 | 4.1 |
| Spectr at detector orth slit A | 0.36 | 0.84 |
| Spectro resolution | | |

**Table 1.** Spectral and spatial optical performances

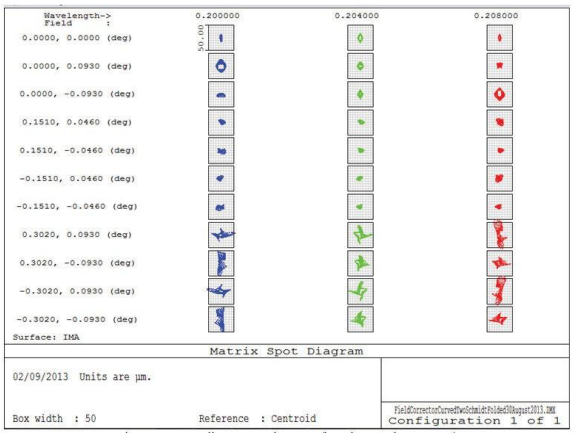

**Fig. 10.** Optimization modeling of FIREBall with Zemax *ray-tracing code*. Spot diagram at detector (box is 50 $\mu$m). Wavelength range is [0.200-0.208 nm]. FOV is [0-0.302 x 0-0.093 deg.]

### 4.2  Elasticity modeling of a freeform reflective diffraction grating

Original active optics techniques were developed by G. Lemaitre [4] to obtain aspherized reflective gratings with rotational symmetry. These instruments basically work at or near normal diffraction angles, with incidence angles of typically 25-30 °. Such ground-based spectrographs were developed and built for CFHT - UV Prime Focus Spectrograph, OHP and PMO - MARLY, OHP - CARELEC, OMP - ISARD, and space-based missions ODIN-OSIRIS-SOHO - CDS and UVCS  [4].

The particularity of this technique is to produce an aspherized grating via two replica stages from a plane passive master. This requires use of an intermediate metallic deformable matrix or submaster. Starting from a plane diffraction grating known as master-grating, the first replication is performed on the plane surface of the unstressed matrix. In a second stage, the grating deformable matrix is aspherized during stressing and replicated on the final slightly concave Zerodur substrate. The radius of curvature of the grating blank is chosen to minimize the thickness of the replica resin layer (Lemaitre et al., 2014)[9] (Fig. 11).

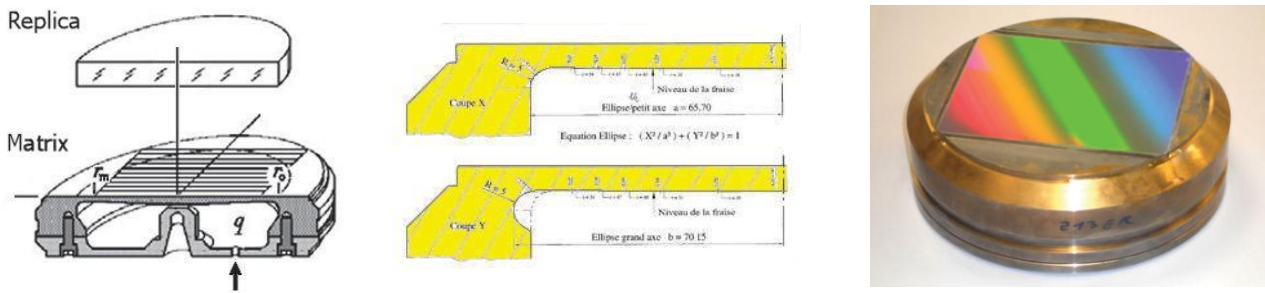

**Fig. 11.** [*Left*] Active optics aspherization of gratings achieved by double replication technique of a metal deformable matrix. A *quasi-constant thickness* active zone is clamped – or *built-in* – to a rigid outer ring which is closed backside for air pressure load control. [*Center*]. For FIREBall, the whole deformable matrix is an axisymmetric piece expect for its inner built-in contour which is *elliptical*. [*Right*] View of matrix with grating before s e c o n d  replication on glass ceramic substrate

**Deformable matrix characteristics :** Principal radii of inner ellipse $a_x$ = 65.70 and $a_y$ = 69.73 mm. Axial thickness $t$ = 6,77 mm. Material AISI stainless steel with $E$ = 201 GPa and $v$ = 0.315. Uniform air pressure $q \mp$ 0.652 $10^5$ Pa. Taking into account a tiny correction of the 5th-order spherical aberration mode the plate were given a slightly thinner axial thickness distribution of amount 0.20 mm from center to edge.

Modeling of the built-in deformable matrix were carried out by Nastran finite element analysis (FEA) code. A convenient number of nodes with adequate boundaries for the built-in conditions provided accurate results with only extremely light modifications – that were not really significant – but prooved FEA modeling results to be as accurate as analytic results.

After realization of the deformable matrix and flat polishing, optical null-test set up allowed to verify the non-axisymmetric part of the freeform shape during uniform loading. Interferograms in Ne-Ne light provide accurate results between theoretical modeling and experiment (Fig. 12).

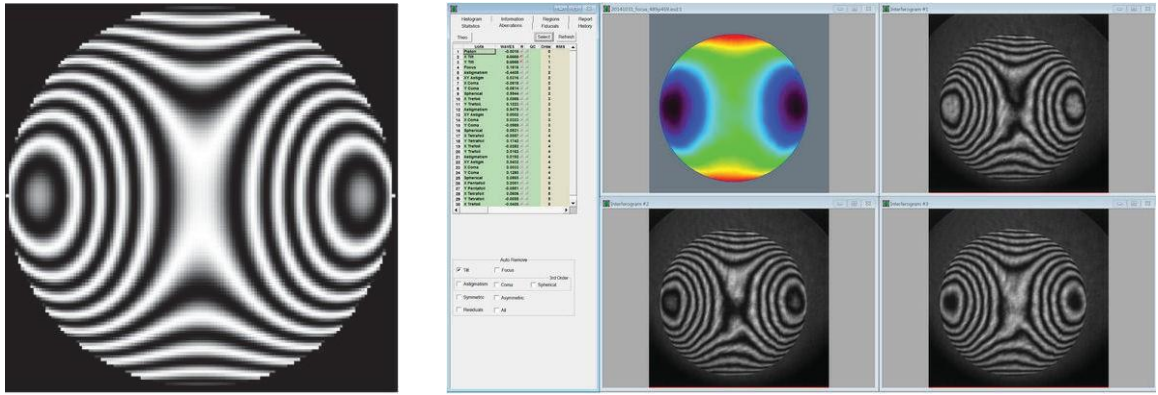

**Fig. 12.** [*Left*] Computed He-Ne interferogram from a null-test singlet-lens compensator. [*Right*] Obtained He-Ne interferogram with the same null-test compensator. NB. : The fringes represent the non-axisymmetric part of the freeform surface   (LAM/AMU)

# 5   MESSIER reflective Schmidt telescope proposal

MESSIER is an experiment proposal named in honor to French astronomer Charles Messier who started compiling in 1774 his famous Messier Catalogue of diffuse non-cometary objects. For instance the Crab Nebulae, named object "M1", is a bright supernova remain recorded by Chinese astronomer in 1054. The present MESSIER proposal is dedicated to the detection of extremely low surface brightness objects. A detailed description of the science objectives and instrument design for this space mission can be found in Valls-Gabaud et al. (2016) [10].

The extremely low surface brightness should reach detections as low as 32 magnitude/arcsec$^2$ in the optical range and 37 magnitude/arcsec$^2$ in the UV (200 nm). Other features require a particular all-reflective optical design as follows,

   1 - a wide field anastigmat telescope with fast f-ratio,
   2 - a distortion-free field of view at least in one direction,
   3 - a curved-field detector,
   4 - an optimal time delay integration by use of drift-scan techniques,
   5 - no spider can be placed in the optical train.

Before the final space proposal, which was restrained to extremely low brightness and ultraviolet imaging and in addition to that of the optical range, our preliminary plan for Messier is to develop and build a 45-50 cm aperture ground-based telescope fulfilling all above features of the optical design. Preliminary proposed designs can be found in Lemaitre et al. (2014) [11] and Muslimov et al. (2017) [6].

## 5.1   Optical design and ray tracing modeling

Our proposed optical design for the ground-based prototype telescope is a three-mirror anastigmat (TMA) with a f-ratio at $f$/2. Compared to a two-element anastigmat design should it be with first element as refractive aspherical plate, where minimization of field aberrations is achieved by:

   → a *balance of the slopes, i.e. balance of the 1st-order derivative* where sphero-chromatism are dominating aberration residuals.

Now, for a first reflective element of a two-element reflective design, the best angular resolution over the field of view is achieved by

   → a *balance of the meridian curvatures, i.e. balance of the 2nd-order derivative* of the aspherical mirror or diffraction grating (Lemaitre (2009)) [4].

The proposed design is made of freeform primary mirror M1, followed by holed flat secondary mirror M2, and then a spherical concave tertiary mirror M3. The center of curvature of M3 is located at the vertex of M1 which is a basic configuration for a reflective Schmidt concept. The unfolded version, with only two mirrors, do not alter anastigmatic image quality (Fig. 13). The optics parameters are optimized for a convex field of view (Table 2). The three-mirror system gives unobstructed access to detector and avoid any spider in the field of view (Fig. 14).

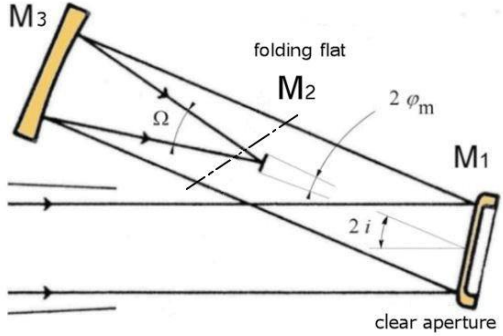

**Fig. 13.** Schematic of Messier reflective Schmidt (M2 holed flat not shown)

**Table 2.** MESSIER optical design parameters.

| Telescope angular resolution | 2 arcsec |
|---|---|
| On-axis circular beam entrance | 356 mm |
| Focal length $f'$ | 890 mm |
| Focal-ratio f/$\Omega$ | f/2.5 |
| Deviation angle 2 $i$ | 22° |
| M1 Elliptic clear aperture $2x_m \times 2y_m$ | 356 × 362.7 |
| Angular FOV $2\varphi_{mx} \times 2\varphi_{my}$ | 1.6°×2.6° |
| Linear FOV | 25×40 mm² |
| M3 mirror curvature radius $R_3$ | −1769 mm |
| UBK7 filter & SiO$_2$ cryostat - Th. | 2 & 5 mm |
| Detector field curvature radius $R_{FOV}$ | −890 mm |

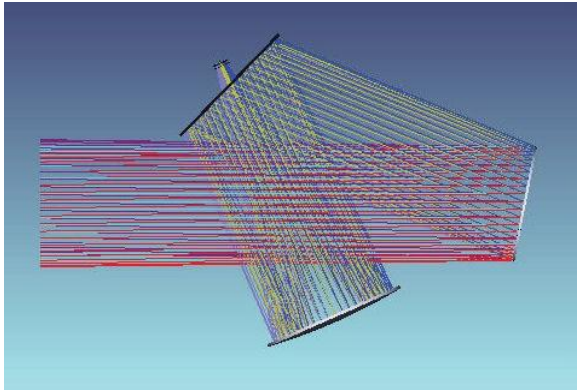
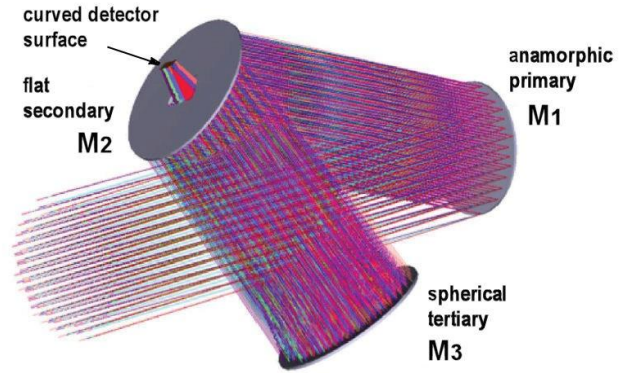

**Fig. 14.** [*Left*] MESSIER layout reflective TMA in the symmetry plane. [*Right*] 3-D view of mirrors and FOV

The primary mirror surface M1 can be defined by the aspheric anamorphic equation as derived from Zemax optics ray-tracing code – which is somewhat similar to Eq.(2) – and denoted

$$Z_{Opt} = \frac{C_x x^2 + C_y y^2}{1+\sqrt{1-(1+K_x)C_x^2 x^2 - (1+K_y)C_y^2 y^2}} + AR[(1-AP)x^2 + (1+AP)y^2]^2, \quad (14)$$

where now (*y, z*) is the symmetry plane of the telescope.
Restraining to the case of a simply bi-curvature surface for the first quadratic term, then setting $K_x = K_y = -1$, denominator reduces to unity. The result from Zemax modeling optimization provides the four coefficients, $C_x = -3.598 \times 10^{-6}$ mm$^{-1}$, $C_y = -3.467 \times 10^{-6}$ mm$^{-1}$, $AR = 2.203 \times 10^{-11}$ mm$^{-3}$, and $AP = -0.01854$. The *total sag of clear aperture* in *x*-direction (i.e. off-symmetry plane) for $x_{max} = 178$ mm is then $Z_{Opt-max} = -35.70\,\mu$m. Zemax iterations lead to RMS residual blur images in agreement with predicted angular resolution (Fig. 15).

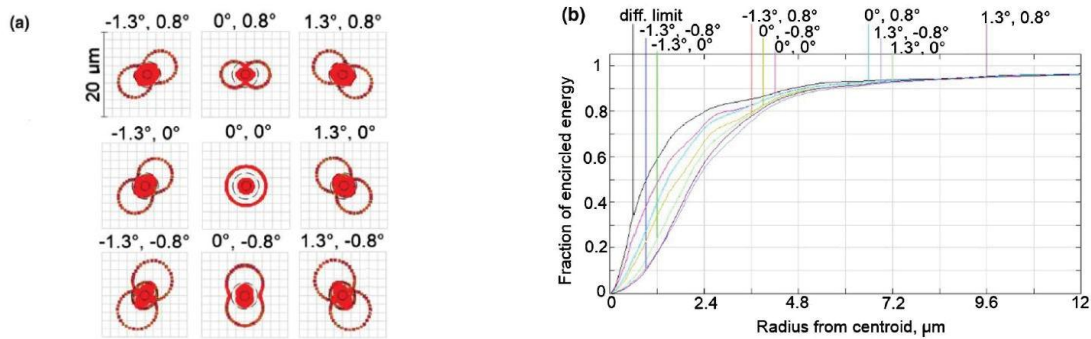

**Fig. 15.** Telescope image quality – curved FOV. **(a)** Spot diagram. **(b)** Diffraction encircled energy

Results from Zemax modeling and spot diagram show that the maximum RMS blur residuals is $\phi_{Zemax}$ = 5.5 µm or $d_{NC}$ = 1.27 arcsec. From the other hand, from Eq. (5) (Lemaitre et al., 2014)[11], a same diagonal semi-FOV (1.52°), and same parameters in Table 2, one also obtains for MESSIER proposal the angular resolution

$$d_{NC} = \frac{3}{256\Omega^3}\left(\frac{3}{2}i + \varphi_m\right)\varphi_m = 1.29 \text{ arcsec.} \quad (15)$$

## 5.2 Elasticity modeling of freeform primary mirror

The non-rotational symmetry of the primary mirror surface refers to an optical surface also called a *freeform surface* (Forbes, 2012)[12], (Hugot et al. 2014)[13]. The present freeform surface for our MESSIER primary mirror is made of homothetic-ellipse level lines. This surface is to be designed through active optics methods (Lemaitre, 1980)[14] where the deformable substrate is aspherized by plane surfacing under stress – also called stress mirror polishing (SFP). The principle uses a uniform load applied and controlled inside a *closed vase form*. The final process delivers the required shape after elastic relaxation of the load (Muslimov et al., 2017) [6].

The *closed vase form* or *closed biplate* is made of twin elliptical vase form blanks in Zerodur assembled together at the end of their outer rings through a layer of 100-150 µm thickness 3M DP490 Epoxy, where elasticity constants are Poisson's ratio $v$ = 0.38 and Young's modulus $E$ = 659 MPa (Nhamoisenu, 2012)[15]. Elasticity constants of the blanks in Zerodur are Poisson's ratio $v$ = 0.243 and Young's modulus $E$ = 90.2 GPa. Modeling with Nastran code led us to make cross optimizations with Zemax code that provided final geometry of the *closed vase form* as follows (Fig.16 and Fig.17). Uniform load of constant pressure $q = 0.687 \times 10^5$ Pa. Inner constant axial thickness $t$ = 18 mm each. Inner axial separation of closed plates 20 mm. Outer thickness of closed form 2 ×28 = 56 mm. Inner radii of elliptic cylinder $2a_x \times 2a_y$ = 356 ×362.66 mm. Cylinder radial thicknesses $t_x \times t_y$ = 18 ×18.33 mm. The length value $\ell$ = 38 mm is the axial distance between middle surfaces of the plates. Inner round corners of radius $R_C$ = 8 mm were adopted. Fig. 18 displays stresses.

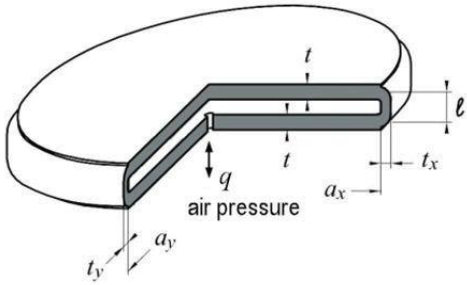

**Fig. 16.** Elasticity design of primary mirror substrate as a *closed vase form* or *closed vase form* made of two identical vase vitro-ceram material linked together with epoxy. The radial thicknesses $(t_x, t_y)$ and height $\ell$ of the outer cylinder provides a semi-built-in boundary which somewhat reduces the size of inner ring radii $(a_x, a_y)$ with respect to that of clear aperture $(x_m, y_m)$
NB.: From anamorphose $x_m/y_m = a_x/a_y = t_x/t_y = \cos i$

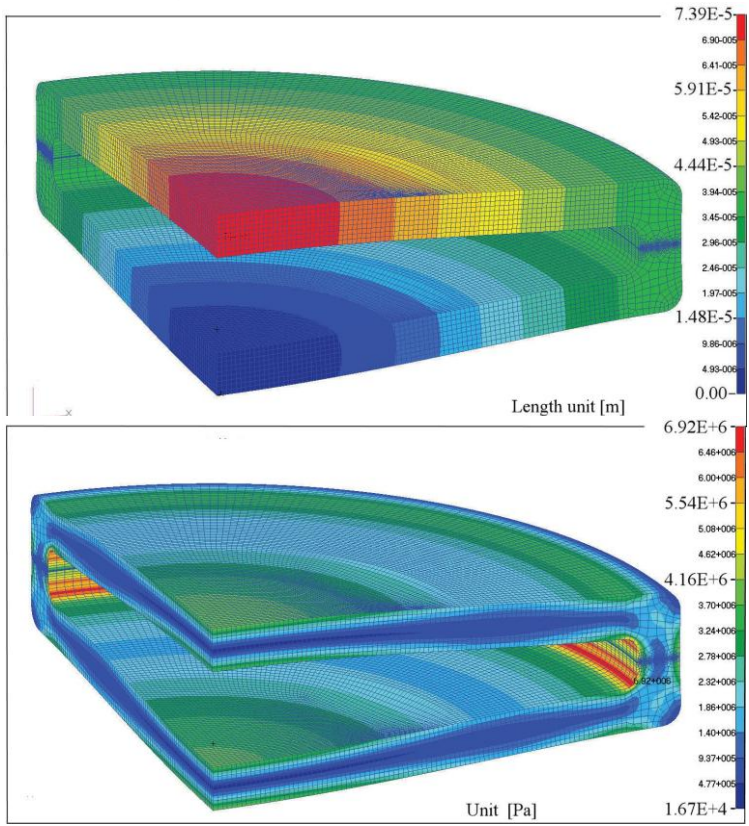

**Fig. 17.** Primary mirror substrate design as a *closed vase form* with FEA Nastran code. All elements are hexahedra. Boundaries are expressed at the origin of displacements $x = y = z = 0$ and freedom along three radial directions in plane $xy$ both at back surface of the figure.
Total axial displacement 69.2 µm

**Fig. 18.** Stress distribution of the *closed vase form* during stress figuring. The maximum tensile stress arises along the internal round corners of the elliptical rings with $\sigma_{max}$ = 6.92 MPa. For Schott-Zerodur this value is much smaller than the ultimate strength $\sigma$ = 51 MPa for a 1-month loading time duration [4].
One may notice that at the symmetry plane the epoxy link reduces $\sigma$ to $\simeq$ 1 MPa

Referring to the law of algebraically *balancing the second-order derivatives* [4] on principal directions of the primary mirror freeform surface – as stated by Eq. (4) – for an elliptical clear aperture where main radii $\rho_m$ are denoted $x$ or $y$ in these directions, we must obtain opposite curvature values at $\rho = 0$ and $\rho = \rho_m$.

One have shown that with a perfect built-in condition – i.e. for a closed vase form with moderate ellipticity and a ring radial thickness, say, at least five times larger than the plate axial thickness – the flexure provides an elliptic *null-power zone* radius $\rho_0$ where the size of the *clear aperture* radius is in the ratio $\rho_0/\rho_m = \sqrt{3/2} \simeq 1.224$.

This means that the useful optical area is convenient for $\rho \in [0, \rho_m]$ but unacceptable for $\rho \in [\rho_m, \rho_0]$.

From our cross optimizations with Zemax and Nastran the ring of the closed vase form provides a noticeable decrease in flexural rigidity, which is equivalent to a semi-built-in condition. Then, compared to a perfect built-in assembly and from our results described in latter subsection, the radii ratio between null-power zone $\rho_0$ and clear aperture $\rho_m$ has become

$$\rho_0/\rho_m = 1.118. \tag{16}$$

This ratio is *significantly* smaller than 1.224. Then from Zemax ray tracing optimizations, as presented in latter subsection, we now take benefit of an important result: the angular resolution is not substantially modified in using the optical surface also up to the inner zone of the ring. The flat deformable surface of the closed vase form requires use of stress figuring by super-polishing, then avoiding any ripple errors of the freeform surface.

→ *Closed vase form* Messier *primary mirror can be aspherized up to its semi-built-in elliptic ring without any lost in optical area and angular resolution.*

Optical testing of the freeform primary mirror must be a precise measurement because of its anamorphic shape. From Eq.(14) the clear aperture this shape presents a total sag of $Z_{Opt-max}$ = -35.70 μm for $x_{max}$ = 178 mm in *x*-direction, i.e. in the off-symmetry telescope plane. Several optical tests could be considered mainly based on a null-test system [6]. These involve for instance lens compensators (Malacara, 2007)[16] and computed-generated holograms (Larionov and Lukin, 2010)[17] and their combinations.

We adopted a singlet lens compensator as component already existing at the lab an providing exactly the correct compensation level of the rotational symmetry mode, that is, 3rd-order spherical aberration compensation of the tertiary mirror. This lens, also called Fizeau or Marioge lens, is plano-convex made in Zerodur-Schott with following parameters : axial thickness $t_L$ = 62 mm, $R_{1L}$ = ∞, $R_{2L}$ = 1180 mm, $D_L$ = 380 mm used on elliptical clear aperture 356×362.7 mm. The axial separation to primary mirror is 10 mm. Remaining aberration is then a balanced anamorphose term to be accurately calibrated by He-Ne interferometry (Fig. 19).

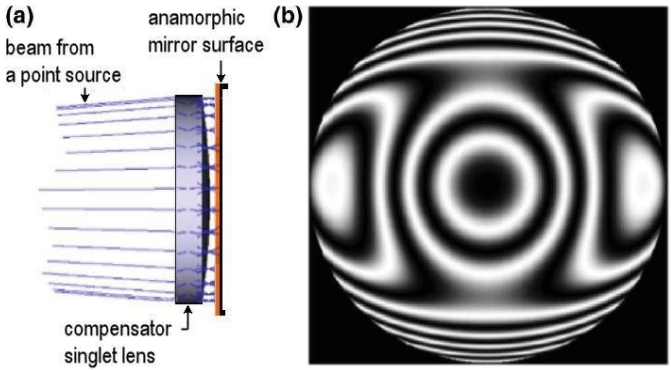

**Fig. 19.** Null-test single lens aberration compensator for the freeform primary mirror. (a) Measurement scheme mounting of the lens with convex surface facing the optical surface. (b) Simulated He-Ne interferogram of remaining fringes to be calibrated (LAM/AMU)

Another important feature of MESSIER telescope proposal is the selection of a *curved detector*, $R_{FOV} = -f \simeq R_3/2$ (cf. Table 2), which allows a *distortion free* design over an active area of 40 × 25 mm². This technology is presently under development by use of either *variable curvature mirrors* (VCMs) or *toroid deformed mirrors* (Lemaitre)[4], (Muslimov et al., 2018)[18]. References can be found on curved detectors in Muslimov's paper [6].

# 6  Conclusion

Freeform optical surfaces generated as a system of homothetic elliptical level lines allow *non-centered optical designs* to provide the best angular resolution. It has been shown that *algebraic balance of local curvatures* –
i.e. balance of second derivatives of the surface – in the principal directions fully minimized the blur residuals images over either a flat or curved field of view.

Modeling of freeform surfaces by development of *active optics techniques* provides extremely smooth surfaces. These surfaces are free from ripple errors as well as for generating the diffraction gratings of FIREBall-MOS balloon experiment by replication gratings of a *deformable matrix*, or for generating *closed vase form* primary mirror by stress polishing of MESIER proposal, which is free from distortion with its curved FoV.

# 7 Annexe - Pupil mirrors

Some particular optical designs may benefit from fruitful optimizations by use of an aspherized pupil mirror – or a pupil transfer mirror that re-image a previous pupil mirror – then providing an efficient field aberration correction.

Referring to Section 2, eq. (4), and to ref. [4][7], it is stated that an aspheric mirror with rotational symmetry – as a two-mirror reflective Schmidt design without tilt of the primary mirror (special case with 100% obstruction design), or a reflective grating working in normal diffraction – is useful if $M_1$ mirror is the pupil mirror and if its *balanced shape* is of the form $z = 3\rho^2 - \rho^4$ in polar coordinates. With normalized aperture radii $\rho \in [0, 1]$ for a full clear aperture mirror, this law states that the local curvatures $d^2z/d\rho^2$ have algebraically opposite values for $\rho = 0$ and $\rho = 1$.

Resuming these general results as a law – **1** axisymmetric case, **2** anamorphic case:

1. *For a* **pupil mirror** *the best balance of an* **axisymmetric** *surface corresponds to the balance of its second derivatives. The local curvatures are opposite at center and edge* $[d^2z/d\rho^2]_{\rho=0} = -[d^2z/d\rho^2]_{\rho=1}$. *A condition that satisfies a null-power radius – in the ratio* $\sqrt{3/2} = 1.225...$ *– located outside the* **circular** *clear aperture.*

2. *For a* **pupil mirror** *the best balance of an* **anamorphic** *surface corresponds to the balance of its second derivatives in main directions. The local curvatures are opposite at center and edge,* $[d^2z/dx^2]_{x=0} = [d^2z/dx^2]_{x=1}$, *... A condition that satisfies null-power radii – in the ratio* $\sqrt{3/2} = 1.225...$ *– located outside the* **elliptical** *clear aperture.*

Above proposition-1 applies to a pupil mirror with a null inclination angle of the principal ray, or an aspheric reflective diffraction grating considered pupil component working at normal diffraction angle. Proposition-2 applies to any pupil mirror or pupil reflective grating used in non-normal incidence angle.

Applied to the optical design for an adaptive anastigmatic five-mirror extremely large telescope, by Delabre (2006, 2008)[19], the ESO project underline the capability to perform *in situ* adaptive optics by use of pupil mirrors $M_4$ and $M_5$. These mirrors allow transferring the pupil of $M_1$ mirror at a location very close to them (Fig. 20). The pupil transfer is a key feature to achieve diffraction limited imaging over a 3-4 arcmin field of view. The 10 arcmin FOV allows space for the implementation of many focal instruments at Nasmyth focii (f/17.5) also with excellent imaging quality.

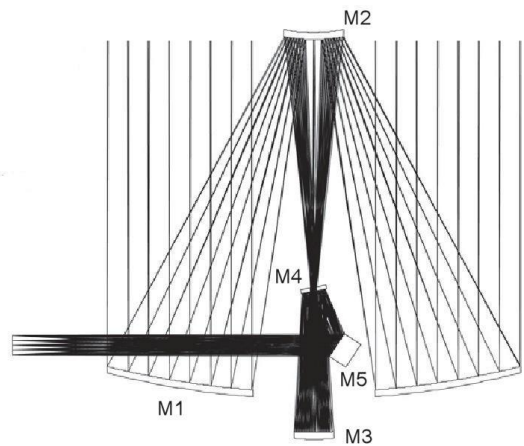

**Fig. 20.** Optical design of the 5-mirror Extremely Large Telescope (ELT/ ESO). The input pupil is the primary mirror $M_1$, segmented aperture diameter 39 m. The pupil transfer is achieved closely to $M_4$ and $M_5$ mirrors which are either side of them. A *field stabilization* (FS) *mirror* and a *multimode deformable mirror* (DM) will be built for the $M_4$ and $M_5$ mirrors or conversely (courtesy ESO)

The shape of $M_1$, $M_2$ and $M_3$ ELT mirrors are an elongated ellipsoid, conicoids and spheroids respectively [4][19], whilst $M_4$ and $M_5$ are flats. However due the very small departure of $M_1$ to that of a paraboloid, this $M_1$ shape has been investigated for optical performance evaluation.

The resulting design with $M_1$ as a paraboloid showed that one of the *pupil transfer mirror*, for instance with a slight asphericity of $M_4$ mirror, provides diffraction limited performance. The image blur is 0.015 arcsec at $\lambda = 1$ $\mu$ – Airy disc diameter 0.012 arcsec – over a field of view of 3 arcmin, The maximum balanced aspheric sag of $M_4$ is one wavelength (630 nm) (Lemaitre, 2006)[20,21]. Ray-tracing modeling with an $M_4$ anamorphic shape could again reduce the blur image size if necessary.

This sag of $M_4$ (i.e. 630 nm PTV departing form plane) is enough small to be fully absorbed by the DM from closed loop control. A major advantage of the design with $M_1$ paraboloid (instead of an elongate ellipsoid) is that the optical testing of each $M_1$ segment (1.4 m in diameter) can be carried by null-test without using corrective holographic plates, i.e. providing a *powerful absolute metrology*.